\documentclass[useAMS,usenatbib]{mn2e}
\usepackage[dvips]{graphicx}
\title[Matching the frequency  spectrum  of PMS stars.]{Matching the frequency  spectrum  of PMS stars by means of standard and rotating models}
\author[Di Criscienzo et al.]{M. Di Criscienzo,$^{1}$ P. Ventura,$^{1}$ F. D' Antona,$^{1}$ M. Marconi,$^{2}$ A. Ruoppo,$^{2}$ 
\newauthor V. Ripepi$^{2}$\\
$^{1}$INAF-Osservatorio Astronomico di Roma, via di Frascati 33, 00040, Monteporzio, Roma, Italy \\
$^{2}$INAF-Osservatorio Astronomico di Capodimonte, via Moiariello 16, 80100, Napoli,
Italy }
\begin{document}

\date{Accepted ???. Received ??? in original form ??}

\pagerange{\pageref{firstpage}--\pageref{lastpage}} \pubyear{2008}

\maketitle

\label{firstpage}

\begin{abstract}
We applied the ATON code to the computation of detailed grids of standard (non--rotating) 
and rotating Pre--Main Sequence (PMS)  models and  computed their adiabatic oscillation spectra,
with the aim of exploring the seismic properties of young stars. 
As, until now, only a few frequencies have 
been determined for $\sim$ 40 PMS stars, the way of approaching the 
interpretation of the oscillations is not unique.
We adopt a method similar to the matching mode method by Guenther and Brown making use, when necessary, also of our rotating evolutionary code to compute the models 
for PMS stars. 
The method is described by a preliminary application to the frequency spectrum of 
two PMS stars (85 and 278) in the young open cluster NGC 6530.
For the Star 85 we confirm, with self--consistent rotating models, previous interpretation 
of the data, attributing three close frequencies to the mode n=4, l=1 and m=0,+1,--1.
For the Star 278 we find a different fit for the frequencies, corresponding to
a model within the original error box of the star, and dispute the possibility that
this star has a T$_{\rm eff}$ much cooler that the red boundary of the radial instability strip.
\end{abstract}


\section{Introduction}
The existence of pulsating stars among intermediate mass Herbig Ae/Be stars 
(1.5 $\la$ M/M$_{\odot}$ $\la$ 4) was originally suggested by \citet{breger1972} 
and confirmed by the pulsation of HR 5999 \citep{kurtz-marang1995} and HD 
104237 \citep{donati1997}. Considering that young PMS stars like the Herbig 
Ae/Be stars evolve through the instability region of post-main sequence stars, 
\cite{marconipalla1998} studied the radial pulsating properties of young 
stars, defining their instability strip (IS) for radial pulsations. Following work 
focused also on the search of further candidates, and today several objects have been 
found and are discussed in the astronomical literature \citep{marconi2004, ripepi2006a, Zwintz2008}.
Most of these stars have more 
than one pulsation frequency, they look like PMS counterparts of the main 
sequence (MS) or early post--MS variables of the $\delta$Scuti type. It is now 
established that the PMS variables can also pulsate in non radial p--modes 
\citep{ripepi2006b,ripepi2007,ZGW2007}. Unfortunately, not many frequencies are identified in 
observations made from the ground, and their analysis then lacks firm 
identification of modes. Nevertheless, new observations are beginning to be 
available and more are expected from the 
program of current (COROT) and up coming (KEPLER) space missions.\\ 
To complicate the things these stars have inherited an angular momentum from the previous protostellar phase, which  explains why most of these stars rotate. 
Recently, the efforts to measure 
stellar rotation periods --- or at least the projected rotational velocities $v 
\sin i$ --- for many stars of different type have  been intensified  \citep{Royer2002a,Royer2002b}. In particular projected rotational velocities in very young cluster stars have been measured \citep{herbst,petit2002} with the aim of providing constraints in the modelling of the evolution of angular momentum during the very early phases of stars formation and PMS stages.
 In particular the study of rotational velocity in the Orion complex \citep{wolff04} has shown that their variation with the evolutionary phase, from the convective PMS to the Main Sequence, is compatible only with decoupling of core and envelope rotation  when the stars begin evolving along the radiative tracks.\\
Since the stellar structure is 
significantly modified by rotation, rotation will affect not only the age and mass 
of the star derived from the location in the HR diagram, but also the 
determination of its physical parameters and of oscillation frequencies in particular.
Till  now the mode 
identification, that is the determination of radial order and spherical 
harmonic  of the observed oscillation spectra, has been based on non rotating 
evolutionary models for the computation of theoretical frequencies, in some 
cases only noticing that the observations present a split, probably due to 
the rotation. As more and more new data from ground and space based 
observations are likely to become available in the near future, we decided to 
begin developing theoretical tools for the interpretation of the PMS 
variability, by using standard and rotating PMS models computed with the code 
ATON \citep{ventura1998,mendes1999}, and deriving their seismic properties by means of 
the adiabatic oscillation code LOSC \citep{scuflaire2007}. In Section 2 we 
discuss our theoretical tools: the input physics of the models, 
and the frequency computation. The method adopted for comparing observed and 
theoretical frequencies is described in Section 3. We test the method, in 
Section 4, on Star 85 of NGC 6530, and compare our results with those 
of \citet{guenther2007}(hereafter G07). We find that  rotating models are consistent with the hypothesis of the split in a 
triplet of the frequency at 180.26 $\mu$Hz. In Section 5, we also discuss the 
comparison with Star 278 of NGC 6530 whereas in Section 6 we summarize the main findings of this paper.

\section{Theoretical tools}
\subsection{The evolutionary code} 
To build the grids of models we have used the  code ATON 
\citep{ventura1998} in its standard version for asteroseismic applications 
\citep{dantona2005} and in its rotational version, including stellar rotation 
according to the formulation  by \citet{endal1976} as described in \cite{mendes1999} 
and in \cite{landin2006}. The program has been recently updated \citep{ventura2007esta}, and we describe in the following its main physical and numerical inputs.

\subsubsection{Input micro-physics} 
The radiative opacities are taken from \citet{iglesias1997}, and  extended by 
the \citet{ferguson2005} tables in the low temperature (T$\le$10000 K) regime;
the conductive opacities, harmonically added to the radiative opacities,
are taken from Poteckhin (2006, see WEB page www.ioffe.rssi.ru/astro/conduct/).
The OPAL equation of state of \citet{rogers1996}, overwritten in the pressure
ionization regime by the EOS by \citet{saumon1995}, is used. The EOS is extended
to the high-density, high-temperature domain according to the treatment by
\citet{stolzmann2000}. A detailed description of the EOS will be presented elsewhere (Ventura $\&$ Mazzitelli, in preparation)

\subsubsection{Convection treatment} 
The convective regions can be described either by the traditional Mixing Length
Theory (MLT) approach \citep{bohm1958}, or following the Full Spectrum of
Turbulence (FST) \citep{canuto1996} prescription. All the models presented in
this paper have been calculated using the FST treatment. Since this work is
focused on PMS evolution, no extra-mixing was considered, though the code presents
the possibility of allowing some extra-mixing (both in the instantaneous and the
diffusive modality) from any convective border.

\subsubsection{Chemical composition}

All the models  were computed with the \citet{grevesse1993} solar 
mixture of heavy elements, and the mass fraction of (X,Y,Z) is (0.70,0.28, 0.02). 
The initial mass fraction of Deuterium is X(D) $\sim$ 2 $\cdot$ $10^{-5}$.

\subsubsection{Rotation}
 Rotation was modelled according to the treatment by \citet{endal1976}.
 This approach accounts only for the hydrostatic effects of rotation,
 neglecting the internal angular momentum redistribution. Three rotation
 schemes are currently implemented in the code, namely i) rigid rotation of the
 whole star, ii) local angular momentum conservation everywhere, iii) local
 angular momentum conservation in the radiative regions and rigid rotation
 of the convective zones. The initial total angular momentum of the star
 is provided as a physical input.
 In this work we used the iii) option in agreement with the analyses of
\citet{wolff04}. The initial angular momentum is chosen to reproduce the
observed surface angular velocity of the star at a given position in the HR diagram.

\subsection{The model grids}
\begin{figure}
\centering
\includegraphics[angle=0,scale=.4]{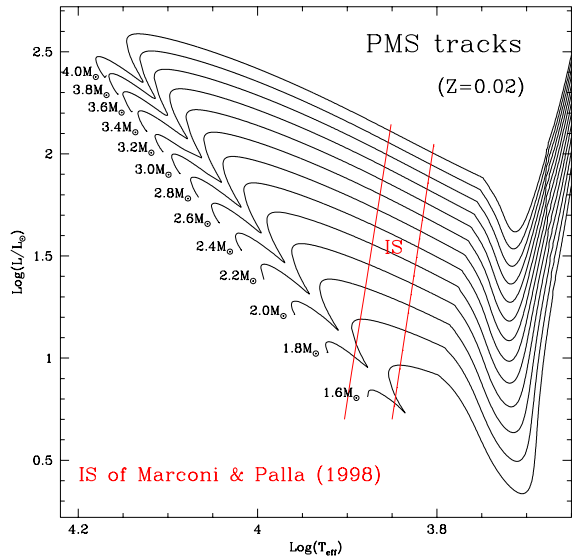} 
\caption{Theoretical HR diagram for the PMS evolution phase. Evolutionary tracks calculated by ATON for the labelled masses are reported. Overplotted is the IS computed by \citet{marconipalla1998}.} 
\label{HRteo}
\end{figure}
Fig. \ref{HRteo} shows a selected set of our  evolutionary tracks from 1.6 to 4M$_\odot$ and the radial IS found by  \citet{marconipalla1998} for the first three radial modes. In detail  it represents the instability strip limits  between the second overtone blue edge  and the fundamental red edge. The red edges for non radial oscillations all lie at T$_{\rm eff }$ larger than this, so the fundamental red edge represents the lower effective temperature for any kind of PMS oscillation \citep{unno1989}.
This limit, however,   depends on the efficiency of convective treatment, being cooler for a smaller efficiency of convection, as the interaction with deep convection quenches pulsations.  The red edge of the radial IS of \citet{marconipalla1998} was obtained with l/H$_p$=1.5 that reproduces the solar radius in their models. Decreasing the mixing lengh parameter from 1.5 to 1, the red edge becomes cooler by 200K (Marconi $\&$ Palla, private comunication).\\
Models were computed from 1.5 to 4  M$_{\odot}$ with a step $\Delta$M=0.02 M$_{\odot}$. The evolution begins in the Hayashi track, and the calculation ends 
when the star has reached the Main-Sequence. 
We memorize the structural quantities needed to perform the computation of the oscillation spectrum. The step in effective temperature of this grid is $\Delta$ T$_{\rm eff}$ $\sim$ 50K.
For rotating models we follow a different approach. As the moment of inertia of 
the model changes along the evolution, together with the total radius and the 
internal mass distribution also the angular and surface velocity change along the 
PMS track, increasing from the youngest model (at the lowest examined 
T$_{\rm eff}$) to the oldest model (at the hottest T$_{\rm eff}$). We build up a sub--grid of rotating models  after having compared observed frequencies with those of the non rotating 
grid and having extracted an initial list of best fit models. \\

\subsection{Adiabatic oscillation code}
\begin{figure*} 
\centering 
\includegraphics[angle=0,scale=.75]{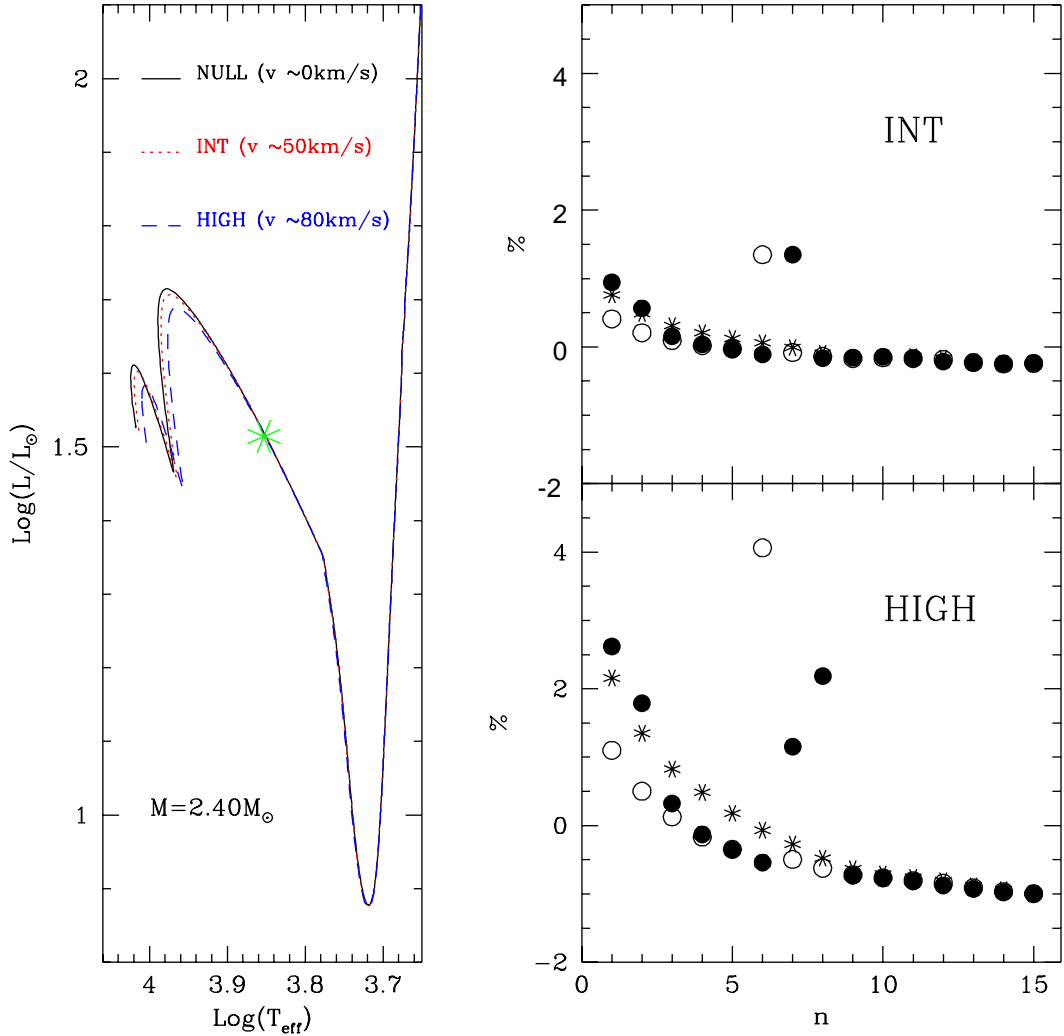} 
\caption{{\it Left panel}: PMS tracks corresponding to M=2.4 M$_{\odot}$ and computed with different rotation velocities (see text). {\it Right panels}: differences (in percentage) for each pair of  n and l(0--asterisk, 1--open dots, 2--filled dots) between frequencies of rotating models and NULL ones for the  models labelled with an asterisk in left panel.} 
\label{ROTteo} 
\end{figure*}
We use the LOSC oscillation code by 
\citet{scuflaire2007} to calculate the frequencies of radial and non radial 
oscillation  modes for each grid point . We limit the spherical degree to l=1,2,3,4 and cover the 
interval in angular frequencies between 0.6 and 60 times the dynamical time\footnote{$\tau _{dyn}$=$\sqrt{\frac{R^3}{GM}}$}. 
All these oscillations are computed with the standard surface boundary condition  $\delta 
P/P+(4+\omega^2)\delta r/r=0$, where $\omega$ = 2$\pi$ f $\tau_{dyn}$ is the dimensionless angular frequency.\\ 
Beyond the radial order $n$ and the degree $l$ another number  characterizes the stellar oscillation frequencies $f_{n,l,m}$: the azimuthal  order  $m$$\in$[-l, l].
The rotation breaks the azimuthal symmetry and removes the 2l+1 degeneracy in $m$.\\
In our modelling, we consider the symmetrical part  of the centrifugal distortion directly  in the computation of rotating stellar models. Then we take into account the first term of perturbation. The perturbation method at the first order\footnote{In fact recently, \citet{Lovekin} using detailed 2D stellar models and a 2D finite difference integration of the linearized pulsation equations to compute non radial oscillations have definitively demonstrated that the eigenfunction can be accurately modelled using  the perturbation theory and a single spherical  harmonic, if rotation is slow enough ($v_{eq} \la$90 km/s).  For these  velocities is justified to neglect perturbation orders greater than one.}  in the rotational velocity $\Omega$ predicts an equidistant frequency splitting $\delta$$_{l,n}$ between consecutive m components within each (l,n) multiplet,i.e: \begin{equation} f_{l,n,m} = f_{l,n,0}+ m \beta _{l,n} \Omega /2\pi \end{equation} where $\beta$, the coefficient of the 
rotational splitting, is calculated directly by LOSC for each model. We stress that, as a novelty in the present computation, $f_{l,n,0}$ is the eigenfrequency of (l,n) mode of the stellar rotating model, structure of which depends  on the rotation rate.\\
In Fig. \ref{ROTteo} we show how these frequencies change for a model of the same mass (=2.40M$_{\odot}$) and effective temperature ($\sim$ 7140K) but with three different values of equatorial velocity of rotation (v=0 (NULL), v $\sim$ 50 km/s (INT) and v$\sim$ 80 km/s (HIGH)); the figure shows the percentage variations of frequencies  with l=0,1,2 and n$\le$26. The general trend is that the differences increase when the rotation velocity increases. It is interesting to note that  for n$>$3-4 the frequencies of the rotating stars  are smaller than those computed for non-rotating models ($<$0.2$\%$ in the INT case and $<$1$\%$ for HIGH rotation velocities). For small $n$, on the contrary, the oscillation frequencies of the rotating models are larger  then the  non rotating ones, up to  2-3$\%$, although the  percentual differences globally decrease for increasing $n$. We see that there is some scatter in the points at low $n$ (see, e.g., the scatter in  dots and triangles at n = 6,7,8). This is  part of the problem  appearing in the computation of frequencies for small $n$ in the LOSC code. The problem will be carefully  investigated in the near future, as PMS stars oscillate mainly at this low n modes.

\section{Comparing observed and predicted frequencies: the method}

The methodology that we will adopt to model the sparse radial and non--radial 
oscillation spectra observed in pulsating PMS stars was originally suggested 
and developed by \citet{guenther2004} and used recently by G07 
to model the oscillations of five pulsating stars in NGC 6530. As recently 
reviewed  by \citet{cunha2007}, there are several other 
methods for mode identification of the frequencies that constitute the observed spectra of a given star and/or to constrain the mass and age of this star. 
These methods use at the same time radial and non radial frequencies. Among 
them, one of the most popular consists in the use of the asymptotic theory and of 
the large and small separation in particular \citep{ruoppo2007}. This 
is a good strategy if the star oscillates in  the asymptotic regime,  or when the 
oscillation data are not well determined but, for the higher quality data 
expected from satellites, e.g. from COROT, a more flexible choice is to use all the 
diagnostic informations obtained from the data, and in particular all the 
discovered frequencies \citep{guenther2004}.\\ 
Indeed, direct fitting of the frequencies is a relatively simple way to find a 
good first solution to the problem of mode identification, and this will become 
an ever and ever better tool with higher quality data. In principle, 
one could proceed with seismic modelling comparing all values for ($l$, $n$) 
to each of the detected  frequencies. The value of degree $l$ is usually limited  (l $\le$ 
4)\footnote{This consideration is based on a geometric statement. In fact in 
distant stars, only the large scale structures can be detected and the 
sensitivity of brightness variations is restricted to modes up to 3 or 4 
node lines at the surface.} and m=0 is assumed when no obvious evidence of 
rotational splitting is present in the observed spectrum. Given 
the $N$ frequencies of a particular star, into our  grid of the 
oscillation spectra we look  for models having $\chi ^2$ $\le$1, where: 
\begin{equation} \chi^2=\frac{1}{N} \sum_{i=1}^N \frac{(\nu_{obs,i}-
\nu_{mod,i})^2}{\sigma_{obs,i}^2+\sigma_{mod,i}^2} 
\end{equation} 
$\nu_{obs,i}$ and $\nu_{mod,i}$ are respectively the observed frequency and the 
corresponding model frequency for the $i$th frequency, $\sigma_{obs,i}^2$ and 
$\sigma_{mod,i}^2$ the associated  uncertainties\footnote{For the 
available observations the model uncertainties are negligible, as they are an order
of magnitude smaller than the observational ones (see Guenher $\&$ Brown, 2004 
for a discussion on this topic)}.
In general the search for the  best--fit  model (which has the spectrum that minimizes the $\chi^2$ ) is done only in a sub-grid made of models  within the 3$\sigma$ error box in T$_{\rm eff}$ and L of the star.
If the program does not find any models with $\chi^2$$\le$1, that means that not all frequencies are matched within the errors, it discards one by one the non-matched frequencies, computes again the $\chi ^2$ for the remaining frequencies  and searches again for models with  $\chi^2$$\le$1. The models with the lower value of the $\chi^2$  is the best-fit model.\\
 When possible, we use the observed frequencies according to the order with which  they are extracted from the Fourier analysis of the data, and we try to fit the frequencies following this order. The order assigns to  the first frequency  the highest probability to be a real one, because it is the first detected by the Fourier analysis and, as a consequence, it deserves the  largest number of options in 
the space of parameters. The same reasoning applies to the following frequencies. Instead, the theoretical frequencies  for each model are examined in order of increasing $l$ (and for each $l$ of increasing $n$) so that the possible matching begins by comparison with radial oscillations first. This approach gives then a higher probability to have a matched frequency with l=0. Afterwords, the search of the closest model frequency for the second one in the list (which, being second has a lower amplitude of oscillation)  is now restricted to model frequencies that have not been already  selected, and so on. \\
 If there is an evidence for the rotation of the star, for example an evident triplet in the observed spectrum, or other rotational indicators from photometry or spectroscopy,  we   calculate a sub-grid of rotating models in  a box around the best fitting model(s) previously obtained. The rotational velocity chosen is that consistent  with  the mean separation observed, with the use of Eq. 1. Then, in the same way, we look again for  the model(s) that minimize  $\chi ^2$   through the models of the new sub--grid.\\
In the next section we apply this method on two test stars.

\section{Example--1: Star 85 of NGC 6530}
\begin{figure} 
\centering 
\includegraphics[angle=0,scale=.44]{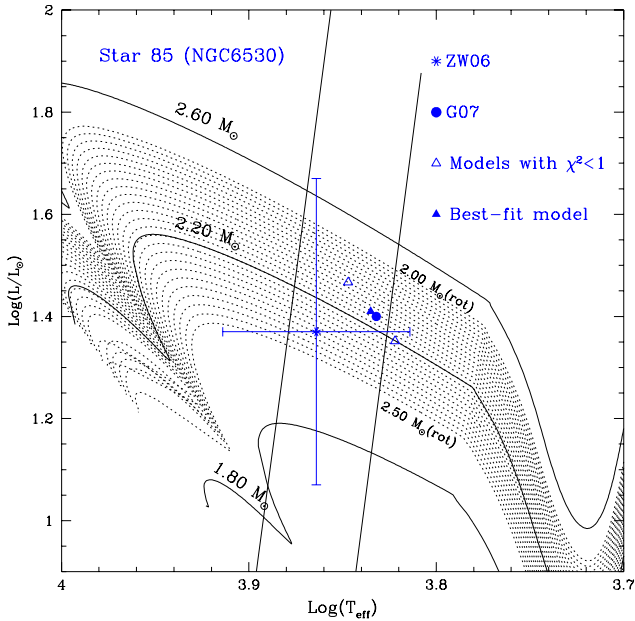} 
\caption{The position of Star 85 in the HR diagram (see text). Solid lines are 
evolutionary tracks calculated with the ATON code with no rotation. Dashed lines 
represent tracks (with mass between 2.00 and 2.50 M$_{\odot}$ and pass in 
mass of 0.02) calculated with option of rotation with an initial angular 
momentum which gives models with v=10km/s near the triangles.} 
\label{HR85} 
\end{figure}
\begin{table}
\begin{center}
\caption{Observed frequencies of Star 85. Those labelled with (I) are found by ZW06 while (II) marks the new frequencies found during the recheck performed by G07; components of the split are labelled with an asterisk. In third, fourth and fifth columns are reported l, n and m of the best--fit model(see text).}
\begin{tabular}{ccccc}
\hline\hline
No. & Frequency ($\mu$Hz)&l&n&m \\
\hline
f1(I)&180.26$^*$& 1&4&0  \\
f2(I)&147.02&0&4&0\\
f3(I)&179.65$^*$&1&4&-1\\
f4(I)&122.58&0&3&0\\
f5(I)&360.48&0&12&0\\
f6(II)&180.95$^*$&1&4&1\\
f7(II)&133.38&2&1&0\\
f8(II)&167.47&2&3&0\\
f9(II)&327.31&2&10&0\\
\hline
\end{tabular}
\end{center}
\end{table}
Among several oscillating PMS stars found in the young cluster NGC 6530 by 
\citet{zwintz2006} (hereafter ZW06), Guenther and collaborators have studied in detail 
the observed spectrum of five stars and were able to constrain the models using 
their mode matching method for several stars. In particular, for Star 85 (WEBDA 
53) several frequencies are available, and they show an 
evident triplet probably due to rotation.
For this star, ZW06 identified five frequencies (labelled with I 
in Table 1). Noticing that two of them  are too much close 
together, G07 have rechecked the data analysis using more recent 
reduction algorithms. In this way they found the lacking 
frequency of  triplet (which appears as  a 1 day$^{-1}$ alias) and three additional frequencies (all labelled in Table 1 with II). 
\begin{figure*} 
\centering 
\includegraphics[angle=0,scale=.8]{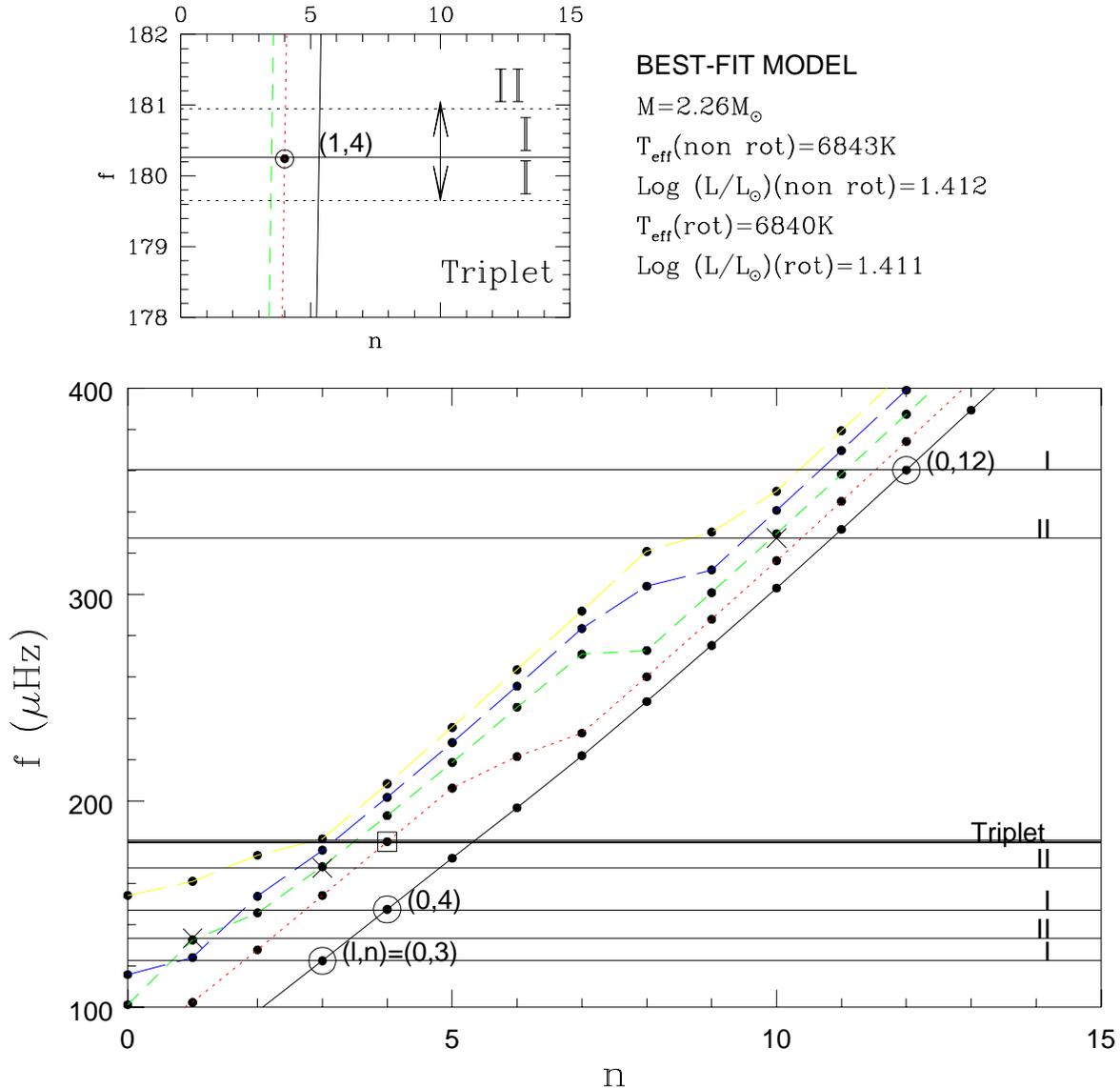}
\caption{Comparison between the observed spectrum of Star 85 and the spectrum of 
the best fit model (filled triangle in Fig. \ref{HR85}). Solid, dotted, dashed and long 
dashed lines connect theoretical frequencies with respectively l=0, 1, 2, 3, 
4. In the small box we report  a zoom around the frequency 180.26 $\mu$Hz in 
order to underline the split do to the rotation of the star.}
\label{F85} 
\end{figure*}
Using $\pm$0.5 $\mu$Hz\footnote{This error correspond to one over the 
time lengh of the  observations.}  as observational uncertainty and keeping only the 
frequencies of the first run (without f3 which belongs to the triplet), G07 
find several models with $\chi^2 \le 1$ , but only one within  their 
observational box around the position of Star 85 in HR diagram and  chose this 
one as the best fit model. Three of the frequencies were fitted with l=0 modes and those 
suspected to be splitted were fitted with an l=1 p mode.
Among the frequencies emerged during the recheck they preserved their best fit 
model and found that the new frequencies lie very close to l=2 p modes, except for 
the highest one (see Fig. \ref{F85} and \ref{ECH85}). 
\subsection{First step: non rotating models}
\begin{figure} 
\centering 
\includegraphics[angle=0,scale=.49]{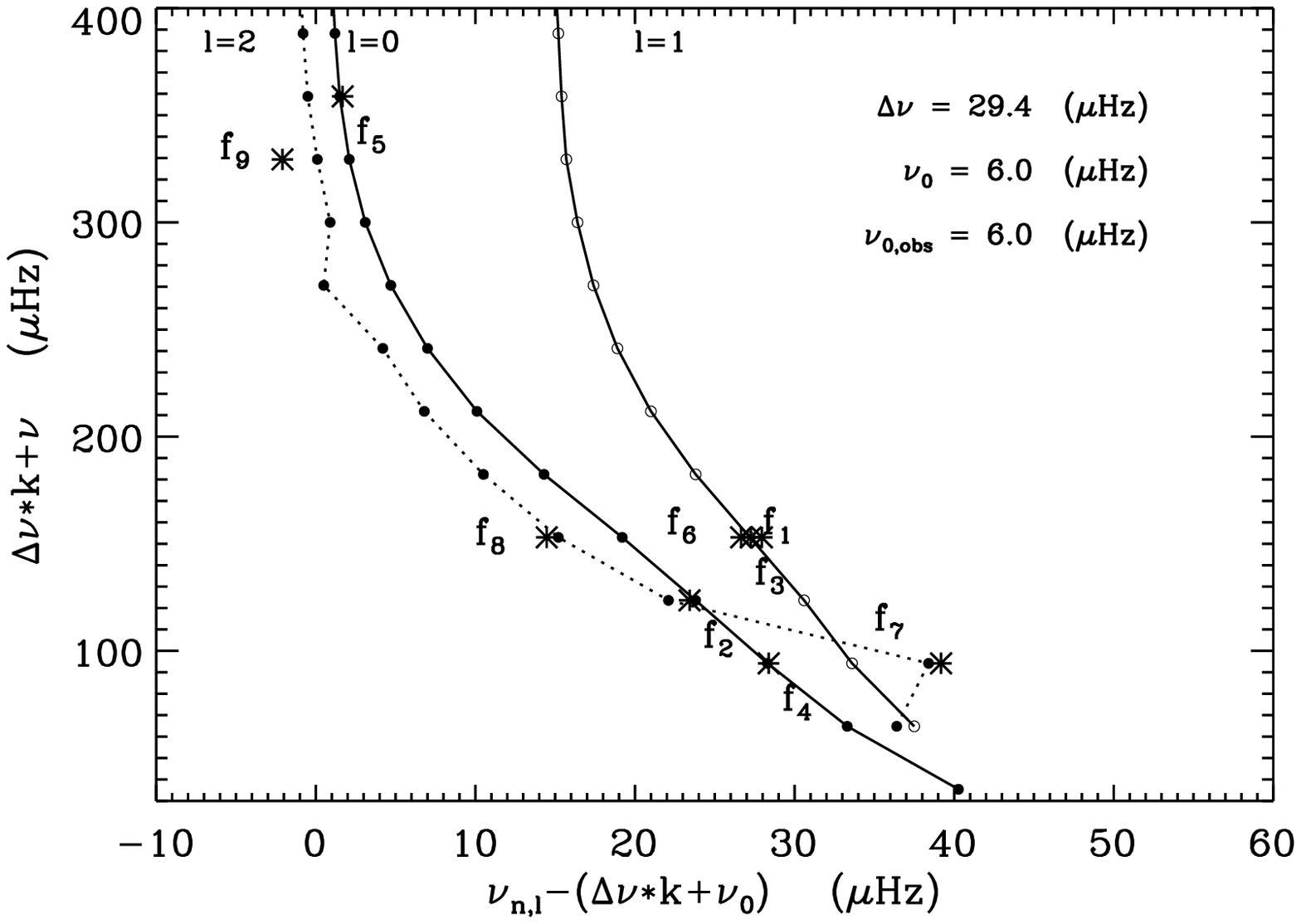}
\caption{Echelle diagram for star 85, using the theoretical frequencies of the best fit model shown in Fig. \ref{F85}. The parameters used to reproduce this plot are also shown.}
\label{ECH85} 
\end{figure}
We have performed a 
search similar to that by  G07, trying to fit the observed frequencies, in the order for which they are 
given in Table 1\footnote{We have chosen  a different order with respect to Table 2 
of G07; in fact for the oscillation modes of the first run we 
have preserved the original order, so the amplitude of oscillations 
decreases going from the first to the last frequency as explained in the previous section. Unfortunately this is possible only for the modes found in the 
first run, for which amplitudes and detail of the Fourier transformations are 
given and not for the other four frequencies which are only provided in 
increasing order in G07.   }. We begin with those found by ZW06, i.e. with those labelled with I in Table 1.
 Assuming for luminosity and effective temperature  of Star 
85 the mean values given in ZW06 we consider an error  of $\pm 
0.05$ in $\log$T$_{\rm eff}$  and $\pm$0.3 in $\log$L/L$_\odot$ (see Fig. \ref{HR85}). Inside this box we search for our  best-fit models. This box is larger than those used by G07 as we have noticed that their best--fit model lies just on the cool edge  of the observational box.
Only three models fit within the errors all four\footnote{We have  discarded f3 which is a splittet frequency.} frequencies 
($\chi^2$ $\le$ 1). In Fig. \ref{HR85} we report these models with triangles; the filled 
one is that with the lowest value of $\chi^2$, i.e. our best fit model which is 
practically overlapped to the model of G07 (filled circle). Also the mode 
identification is the same, with the three frequencies f2, f4, f5 matched with 
(l,n)=(0,4), (0,3), (0,12), and the splitted one (f1)  with (l,n)=(1,4) (see Fig. \ref{F85}). \\
Alternatively, by fixing these four pairs of (l,n) to the observed frequencies,  the models  with the lowest $\chi^2$ (including the three ones previously found) all lie on a diagonal of HR diagram (see Fig. \ref{HR85}) as found for the case B by G07. 
This  can be understood in the following way: by choosing these pairs (l,n) we assign a large separation ($=f_{n,l}- f_{n-1,l}$), which in asymptotic regime depends linearly on luminosity and temperature as shown by  \citet{ruoppo2007}.\\
Concluding, using similar hypotheses, we obtain the same result of G07, i.e. the same best-fit model. The same 
models are selected, when we search among  all the grid models.
Adding the frequencies found during the recheck of the data doesn't 
change the model with lowest $\chi^2$ but in this case the lowest $\chi^2$ is not 
$\le$1. Looking in detail to the comparison, the main reason for this is that the three new frequencies (f7, f8 and f9 in particular) are not perfectly matched ($\Delta f > \sigma_{obs}$ from the nearest ones with l=2). This problem was also 
found in the G07 investigation. A possible interpretation is 
given in the next subsection. 
\subsection{The second step: rotating models}
Now we implement the comparison with rotating evolutionary 
models. Using the observed  splitting $\Delta$f=0.643$\pm$0.09  $\mu$Hz (=$\frac{f6-f3}{2}$), Eq. 1 and the radius of the best--fit model found in the previous section we derive a rotation rate of  v$_{\rm eq} \sim$ 10km/s.
We  calculated a  
sub-grid of rotating models choosing an initial angular momentum  that allows to obtain the 
predicted velocity near the position in HR diagram of the best-fit model 
previously found.\\
The model grid  is restricted to  2$\le$ $\frac{\rm M}{\rm M_{\odot}}$ $\le$2.5 and has the same step in T$_{\rm eff}$ and mass of the non--rotational grid. The rotational velocity is v$_{\rm eq}$ $\sim$ 10 km/s around log T$_{\rm eff}$=3.83 .\\
We compared again observed 
and calculated spectra looking for a good match in terms of $\chi^2$. 
Such a slow rotation velocity modifies very little both the structure of the star and its position in HR diagram, so we obtain identical results. 
Although we find  the same three best 
fit models by considering the fit of the first frequencies only, when we add the three additional frequencies the  rotating models  match them better but continues to lie $ 1 \sigma$  away  from the nearest theoretical frequency. Maybe it would be useful to recheck  the data of the second round. 
An alternative explanation for this discrepancy is the following. \citet{das2002} tested  the results of non adiabatic models on the $\beta$ Cep stars. One of the effects  of an even slow rotation, is 
to influence  mode properties, when there is a small frequency distance between 
modes  with the same azimuthal  order m and spherical harmonic degree l 
differing by 2. In this case each of the coupled modes must be represented by a 
certain superposition of spherical harmonics of the two models involved. This 
effect was invoked to explain the nonradial character of pulsation in 
$\beta$ Cep stars but it could be a possible motivation for the worse match of the three last frequencies, either with the closest  frequencies with l=2, or with those with l=0. It is important to stress that this is only a 
suggestion and that a definitive statement will be possible only when better data will be available.

\section{Example--2: Star 278 of NGC 6530}
\begin{figure} 
\centering 
\includegraphics[angle=0,scale=.42]{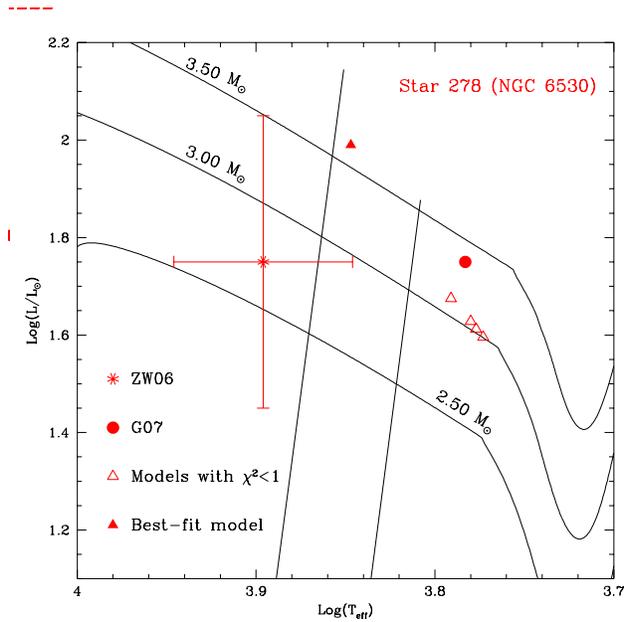} 
\caption{The position of Star 278 in the HR diagram (see text). Solid lines are evolutionary tracks calculated with the ATON code with no rotation. } 
\label{HR278} 
\end{figure}
\begin{table}
\begin{center}
\caption{Observed frequencies of Star 278. In the third and fourth column we 
have reported the l and n identification of the best fit model obtained searching 
inside the observational box. With asterisks are labelled those frequencies 
which are not matched. For this frequencies we report l and n of the nearest 
frequency ($\Delta$f2 $\sim$ 1.3 $\mu$Hz and $\Delta$f7 $\sim$ 0.8 $\mu$Hz)}
\begin{tabular}{cccc}
\hline\hline
No. & Frequency &l&n\\
\hline
f1&  83.3&2&3\\
f2& 140.3&0*&9*\\
f3& 187.7&3&11\\
f4&  48.3&0&2\\
f5& 109.8&2&6\\
f6&  69.6&4&3\\
f7& 138.7&2*&8*\\
f8& 181.5&2&11\\
f9& 160.8&1&10\\
\hline
\end{tabular}
\end{center}
\end{table}
Now we analyze Star 278 in NGC 6530. Among the nine frequencies emerged from the Fourier analysis of the data performed by 
ZW07, G07 were able to fit all  with l=0,1,2 p modes, except the lowest one. 
All their best fit models lie at a much lower T$_{\rm eff}$ (1800K below the mean temperature given by ZW06). The 
authors support this interpretation considering that this star, together with 
some other  stars in the cluster, might have been  dereddened in excess in ZW06, when their  color 
temperatures were derived. To support this hypothesis, they  notice  that the amount of 
gas near these stars correlates linearly with the difference between model and 
observed surface temperature.\\ 
This T$_{\rm eff}$ is really very  low, if compared to the red edge of the IS calculated by \citet{marconipalla1998} ($\sim$ 600K see Fig. \ref{HR278}).
As pointed out  in Section 3  the cool edge of the radial IS also applies to non radial oscillations, and it only depends on the efficiency of convective treatment, being cooler for a smaller efficiency of convection.
In particular, the red edge of the radial IS of \citet{marconipalla1998} was obtained with a value of l/H$_p$=1.5 and its effective temperature decreases by $\sim$ 200 K for l/H$_p$=1 remaining still hotter than the best fit of G07.
The efficiency of convection in pre-MS can be tested for the convective Hayashi phase of the low mass stars by using the location in the HR diagram of pre--MS binaries. The binaries however do not provide a calibration for the stars located on  the 'radiative' phase of the tracks, which is practically independent on the convection model. In fact, the analysis of dynamical PMS masses shows a good agreement with all stellar models for M$\ge$1.2M$_{\odot}$ (Hillenbrand $\&$ White 2004) for which there are no stars on the convective tracks in the examined sample. We  know that at the solar T$_{\rm eff}$ (5780K) the l/H$_p$ ratio must be quite large for models to fit the solar radius. We also have further hints for models at larger T$_{\rm eff}$, where the convective layer becomes very thin. Non local models of convection (Kupka $\&$ Montgomery 2002) for A stars between 7200 and 8500 K can not be easily compared to MLT results, but comparable values of the maximum convective flux in the H--convection zone are obtained with $\alpha$=1 at T$_{\rm eff}$ $\sim$ 7100K (log T$_{\rm eff}$=3.85) and with $\alpha$=0.4 at T$_{\rm eff}$ $\sim$ 8000K (log T$_{\rm eff}$=3.90).  Thus non local models shows that the convection efficiency is much smaller than in standard MLT models at the blue boundary of the IS, but rapidally increase while convection deepens. It is clear that a further exam of the IS boundaries is needed. For the present test we prefer to  search in the observational box allowing only temperatures 
hotter than the IS red edge obtained by \citet{marconipalla1998}.\\ 
We have studied this star using our grid of non rotating models since no 
evidence of relevant rotation is observed in the spectrum. This does not  mean that the star does not rotate, 
but it could mean that the star rotates so slow that the effect on the frequencies 
is smaller than the observational uncertainties. Using the 
luminosity and temperature given by ZW06 and the same size of the observational box used for 
Star 85 we are able to reproduce only seven of the nine frequencies. It is 
important to say that among the models able to reproduce seven frequencies only one model in the grid has $\chi^2$ $<$1: the 
position of this best fit model in the HR diagram is reported in Fig. \ref{HR278} and the mode 
identification in Table 2. The only non matched 
frequencies are those two which lie very close to each other.
As we noticed discussing the results for Star 85 also in this case the theoretical frequencies
which are close  to these two observed ones  have $l$ 
that differs by 2 (see rows marked with asterisks in Table 2) and   they may emerge in the observed spectra  as coupled modes due to the rotation, even  if slow enough that we 
don't see it as split of frequencies with l $\neq$ 0, as suggested \citet{das2002}. 
But a definitive conclusion can be done only when new data will be available. In this direction also goes the observation that the two frequencies that we  are not able to match with the theoretical ones are so close, suggesting the presence of a third one, that could be a part of a triplet due to the rotation. Unfortunately, is impossible to find this frequency from the available data.
\section{Conclusions}
 In this paper we have discussed the potentialities  of a method built up for the interpretation  of the observed multi-frequency spectra of pulsating stars, giving us the possibility of constraining its mass and age.\\
The method  relies on the direct comparison of the observed oscillation spectrum  of a given star with a database of theoretical ones. How well a given theoretical spectrum matches with the observed one is measured by $\chi^2$. Each frequency set in the database was computed using the LOSC adiabatic oscillation code to derive the pulsation frequencies of the  evolutionary models computed with the ATON code. The characteristics of the grid were  reported in Section 3 and it is fine enough to determine a  best fit model to the oscillation data as constrained solely by observed frequencies.
Till now, the method has been applied only to the interpretation of the PMS pulsating stars, but the intention is to extend it also to later evolutionary phases.\\
We have included the rotation of the star not only  considering the removal of the $m$-degeneracy, but also in the calculation  of the evolutionary models  from which we compute the pulsational frequencies. During the evolution, the angular momentum distribution assumes rigid rotation in the convective layers, and conservation of the momentum shell by shell, in the radiative core, as suggested by \citet{wolff04}. We have shown that, for the some physical inputs, the variations of frequencies with rotation are small but significative expecially when better data will come, from current space missions for example. In particular, our calculations have demonstrated that  the differences increase when the rotation velocity increases. While  the frequencies of the rotating stars  are smaller than those computed for non-rotating models for  n$>$5-6, inversely for small $n$ the oscillation frequencies of the rotating models are larger  then the  non rotating ones, up to  2-3$\%$ for stars that rotate with v$_{\rm eq}$ $\sim$ 80 km/s.\\  
We have tested the mode-matching method on data of two stars of NGC 6530.
For the Star 85 we confirm with self--consistent rotating models the interpretation 
of the data performed by G07, attributing three close frequencies to the mode n=4, l=1 and m=0, +1, --1. We also found that the remaining three frequencies are better matched when rotating models are used.
For the Star 278 we find a different fit for the frequencies, corresponding to
a model within the original error box of the star, and dispute the possibility that
this star has a  T$_{\rm eff}$ much cooler that the red boundary of the radial instability strip  published by Marconi $\&$ Palla (1998).\\
\section*{Acknowledgments}
It is a pleasure to thank R. Scuflaire and J. Montalban for having provided us the  Li\`ege OScillation Code (LOSC) and the referee W. W. Weiss for his useful suggestions that improved the final version of the paper. \\
Financial support for this study was provided by MIUR under the PRIN project ``Asteroseismology: a necessary tool for the advancement in the study of stellar structure, dynamics and evolution'', P.I. L. Patern\'o.\\

\end{document}